\documentclass[sigconf]{acmart}
\acmConference[ESEC/FSE 2022]{The 30th ACM Joint European Software Engineering Conference and Symposium on the Foundations of Software Engineering}{14 - 18 November, 2022}{Singapore}
\AtBeginDocument{%
  \providecommand\BibTeX{{%
    \normalfont B\kern-0.5em{\scshape i\kern-0.25em b}\kern-0.8em\TeX}}}

\newcommand{\sectopic}[1]{\vspace{0.2em}\par\noindent{\textit{\bfseries #1}}}

\newcommand\addauthornote[1]{%
  \if@ACM@anonymous\else
    \g@addto@macro\addresses{\@addauthornotemark{#1}}%
  \fi}

\begin{document}

\title{COREQQA - A COmpliance REQuirements Understanding using Question Answering Tool}

\author{Sallam Abualhaija}
\email{sallam.abualhaija@uni.lu}
\affiliation{%
    \institution{University of Luxembourg} 
  \country{Luxembourg}
}

\author{Chetan Arora}
\email{chetan.arora@deakin.edu.au}
\affiliation{%
    \institution{Deakin University} 
  \country{Australia}
}

\author{Lionel Briand}
\email{lionel.briand@uni.lu}
\affiliation{%
    \institution{University of Luxembourg}
  \country{Luxembourg}
}
\affiliation{%
    \institution{\& University of Ottawa}
  \country{Canada}
}

\renewcommand{\shortauthors}{Abualhaija, et al.}


\begin{abstract}
  We introduce COREQQA, a tool for assisting requirements engineers in   acquiring a better understanding of compliance requirements by means of automated Question Answering. 
  Extracting compliance-related requirements by manually navigating through a legal document is both time-consuming and error-prone. COREQQA enables requirements engineers to pose questions in natural language about a compliance-related topic given some legal document, e.g., asking about \textit{data breach}. 
  The tool then automatically navigates through the legal document and returns to the requirements engineer a list of text passages containing the possible answers to the input question. For better readability, the tool also highlights the likely answers in these passages. 
  The engineer can then use this output for specifying compliance requirements.
  COREQQA is developed using advanced large-scale language models from BERT's family. COREQQA has been evaluated on four legal documents. The results of this evaluation are briefly presented in the paper.
  The tool is publicly available on Zenodo~\cite{abualhaija_sallam_2022_6653514}.
\end{abstract}


\maketitle

\section{Introduction}~\label{sec:introduction}

With the growing reliance on personal data and confidential information, software systems are increasingly subject to compliance against regulations to enforce necessary safeguards for information protection and human safety~\cite{Janssen:2020,Leidner:21}. Failing to comply with relevant regulations can lead to legal, fiscal or reputational implications for an organization. Regulatory compliance is regarded as an essential yet challenging task by the Requirements Engineering (RE) community~\cite{Otto:07,Berenbach:09,Sleimi:18}.

\begin{figure}
\centering
  \includegraphics[width=0.5\textwidth]{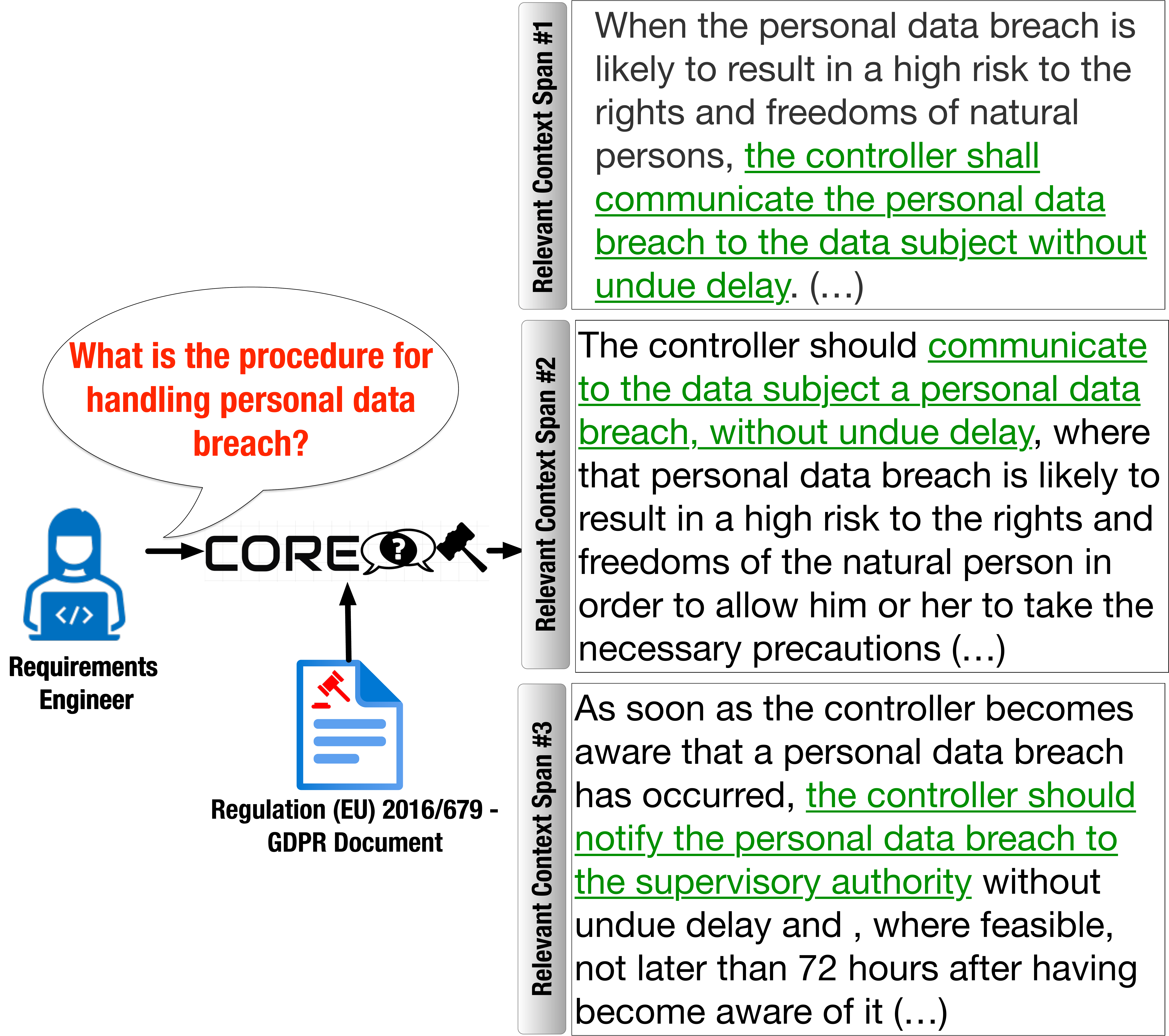}
    \vspace*{-1em}
  \caption{{Example of COREQQA's QA assistance on GDPR~\cite{GDPR}.}}
  \label{fig:example}
  \vspace*{-1em}
\end{figure}

In this paper, we propose the tool COREQQA (\textbf{Co}mpliance \textbf{Req}uirements Understanding using \textbf{Q}uestion \textbf{A}nswering). COREQQA is motivated by actual practical needs, considering that requirements engineers are being increasingly involved in software compliance against relevant regulations, e.g., all software systems in Europe must comply with GDPR (Regulation (EU) 2016/679) -- the European regulation on data protection, privacy and personal data transfer~\cite{GDPR}. Manually handling compliance requirements is tedious and error-prone since requirements engineers have to read through entire legal documents. Such documents are usually hefty, contain complicated natural language (NL) structures, frequently refer to external regulations, and are not easy to peruse without legal expertise~\cite{Sleimi:19,Abualhaija:2022}. 

An automated Question Answering (QA) tool such as COREQQA helps requirements engineers efficiently navigate through the compliance-related content of legal documents.
QA is the task of automatically finding the answer to a question posed in NL from a given text passage. In our work, we refer to a single text passage as a \textit{context span}. 
Instead of reviewing long, complex legal documents, COREQQA enables requirements engineers to ask a question about a compliance-related topic, and then returns a list of relevant context spans in which the answer is likely to be found. This way, COREQQA pinpoints the requirements engineers to the portions of the legal document where they need to invest their efforts and time. 

We illustrate in Figure~\ref{fig:example}, the QA assistance provided by COREQQA to a requirements engineer, who is interested in understanding the regulations related to personal data breach. 
The example question is specifically related to the process for handling personal data breaches. The answer to this question is mined in the GDPR text. 
The legal obligations with regard to handling data breaches can have a significant impact on the software development process, e.g., sending notifications to different responsible agents within legally-binding time constraints.
COREQQA assists the requirements engineer in retrieving relevant information to define the respective compliance requirements for handling data breaches for the software system under development. 
As we elaborate in Section~\ref{sec:architecture}, COREQQA returns as output the top-$N$\footnote{$N$ is a configurable parameter and is set $N=3$ for the example question in the figure} relevant context spans from a given legal document and the potential answers highlighted in the context spans. 
COREQQA builds on large-scale natural language processing (NLP) language models for solving the QA task (also widely known as \emph{machine reading comprehension (MRC) task}~\cite{Jurafsky:20}). QA models for MRC generally assume that for each question, the relevant context span (containing the answer) is known a priori. 
Developing a practical QA tool with this restriction is infeasible, as requirements engineers have no means of knowing the exact context span with the correct answer in advance. Therefore, in COREQAA we first find top-$N$ relevant context spans which likely contain the answer. To do so, we compute the semantic similarity between each context span in the legal document and the input question. Then, we demarcate the answer to the input question using the QA models.

We further observe that information relevant to answering the question could be found in multiple non-contiguous context spans (i.e., different sections in the same legal document). For example, the top-3 context spans selected in Figure~\ref{fig:example} are all directly relevant for answering the question. The first two spans (retrieved from different sections of the document)
explain the process of communicating breach details to the data subject, while the third span
specifies how to communicate breach details to the supervisory authorities. Thus, by retrieving multiple relevant spans and further highlighting the likely answers, COREQQA enables the requirements engineer to specify a complete and precise set of compliance requirements. 
We leave configuring the $N$ parameter to the requirements engineer. While selecting higher values of $N$ entails more time and effort for reviewing the retrieved context spans, we believe that using COREQQA is still much more cost-effective in practice than manually traversing the entire legal document for the answer. 

In the remainder of this tool demonstration paper, we elaborate the architecture of the tool, the dataset that we generated for developing COREQQA as well as an end-to-end usage scenario.

\section{Tool Architecture}~\label{sec:architecture}

\begin{figure*}
\centering
  \includegraphics[width=\textwidth]{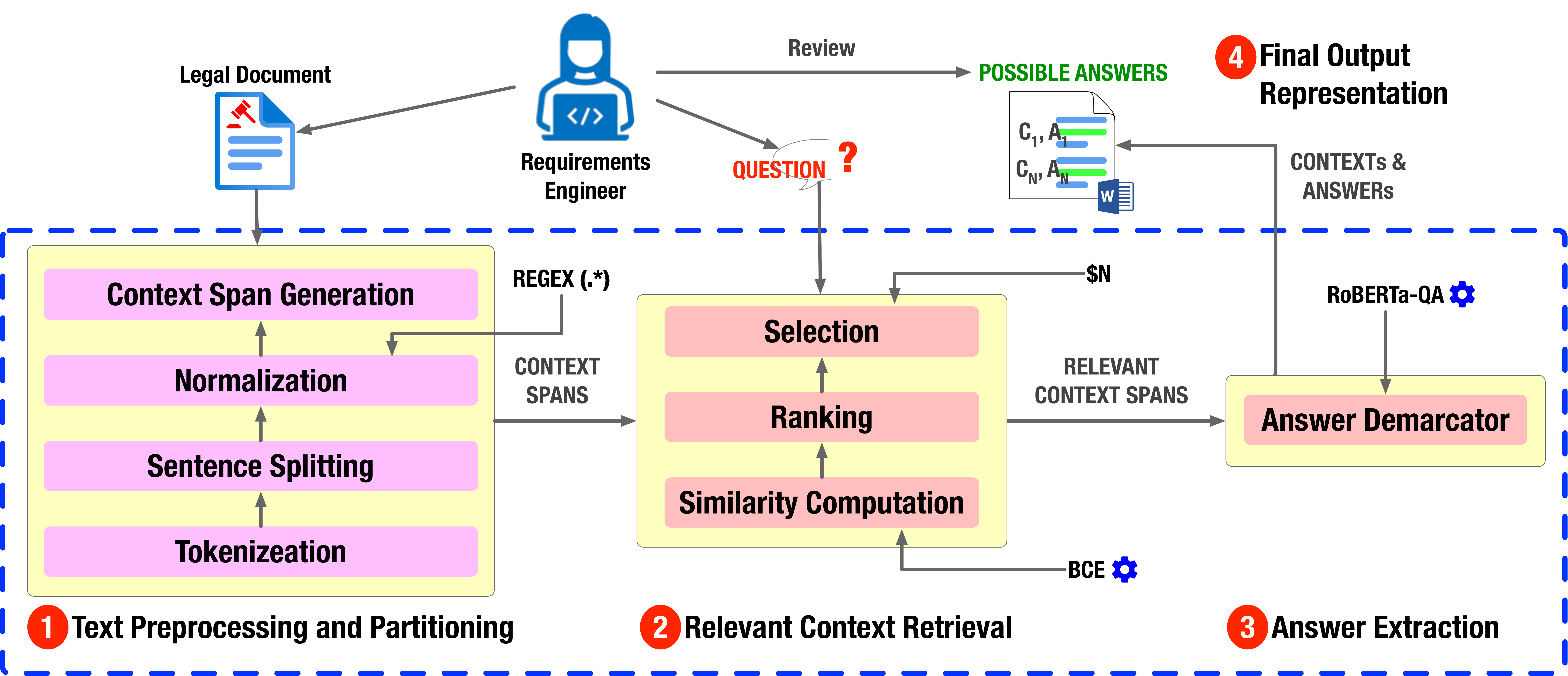}
  \caption{Overview of the end-to-end architecture of COREQQA.}
  \label{fig:tool}
\end{figure*}

The end-to-end architecture of COREQQA is illustrated in Figure~\ref{fig:tool}. 
COREQQA aims at answering a given question posed by a requirements engineer in NL on some legal document. Below, we elaborate the main steps of the tool marked as 1 -- 3 in Figure~\ref{fig:tool}. We implemented COREQQA in Python 3.8. 

\subsection{Text Preprocessing}~\label{subsec:preprocessing}
In the first step, COREQQA parses the legal document and applies a simple NLP pipeline which is composed of tokenization and sentence splitting. 
The tool then applies a set of regular expressions for normalizing the text (e.g., removing periods from the ending of acronyms, ``Art.'' becomes ``Art''). The motivation for normalizing the text is to improve the accuracy of sentence splitting. We operationalize the NLP pipeline using NLTK library~\cite{Bird:09,NLTK}, and the \textit{re module} in Python for regular expressions\footnote{\url{https://docs/python.org/3.8/library/re/}}. 
In this step, we further partition the legal document into context spans. Due to technical constraints of underlying QA models, the maximum size of each context span is 512 tokens. To maintain coherence, we split the document into paragraphs first, and then check their size. Each paragraph fitting this size limitation is regarded as one context span. Otherwise, we split the paragraph into half, and check the size again. This process is iteratively performed until size limitations are met.  
The output of this step is the list of context spans representing the input legal document.

\subsection{Relevant Context Retrieval}~\label{subsec:contextRetrieval}
In the second step, COREQQA computes the semantic similarity between the input question and each context span generated from the previous step. 
We implement this step using the BERT cross-encoder (BCE) model available in the Sentence-Transformer 2.1.0~\cite{Reimers:19} provided by Hugging Face\footnote{\url{https://huggingface.co/}}.  
BCE takes as input two text fragments, and returns as output a score between 0 and 1 indicating how semantically similar the two fragments are, with 1 being identical. To assess the relevance of the context span, we first compute BCE between the question and each sentence in the context span and then assign to the context span the maximum score achieved by any sentence. The intuition behind this computation strategy is that only a portion of the context span is expected to contain the likely answer to the input question. 

Once we compute a score per context span, we rank the spans in descending order. We do this using the sort function from \textit{pandas} in Python\footnote{\url{https://pandas.pydata.org}}. 
The result of this step is a list of top-$N$ relevant context spans to the input question. We keep the value \textit{N} as a parameter that can be initialized by the requirements engineer.
The value of $N$ depends on the practical context in which COREQQA is applied. Selecting large values (e.g., $\geq 10$) entails that the requirements engineer will review many 
context spans to gain a better understanding of the compliance requirements associated with the question. Selecting small $N$ values (e.g., $\leq 3$) entails less context spans to be reviewed, but also a higher risk for the right answer not to be found in any of the top-$N$ relevant spans. 
In Figure~\ref{fig:example}, we show an example of top-3 context spans retrieved in this step. The default value of the configurable parameter $N$ is set to 5 in COREQQA, based on previous experiments~\cite{Abualhaija:2022}.

\subsection{Answer Extraction}~\label{subsec:answerExtraction}
In the last step, we pass on the top-$N$ context spans deemed relevant in the previous step together with the input question to a pre-trained QA model. In our work, we apply the Transformers library to extract answers for the given question using the RoBERTa QA model~({\emph{roberta-base-squad2-distilled}}). RoBERTa then extracts from each context span a potential answer for the input question. 
In Figure~\ref{fig:example}, we highlight the extracted answers in green. 
We note that the engineer has access to the context spans already in the previous step. Thus, this step is not essential for providing assistance to the requirements engineer in understanding the questions related to compliance requirements. However, highlighting the answer in the context span improves readability and leads to a more efficient reviewing process. In practice, when the engineer selects a larger number $N$ (say 10), it is then advantageous to demarcate the answer automatically to help the engineer quickly navigate through the context spans. 

\subsection{Final Output Representation} 
For presenting the final output of the tool to the requirements engineer, we export for each question the top-$N$ context spans and the highlighted answers within these context spans as a Microsoft Word document. The document also shows for each context span a confidence score (range 0--1) of the answer highlighted in the span. This confidence score is automatically assigned to  the extracted answer by the QA model. 
We use the python-docx library v0.8.11~(\url{https://python-docx.readthedocs.io}) for exporting the output and visualizing highlighted answers in the relevant top-$N$ context spans.

\section{Evaluation}~\label{sec:evaluation}
COREQQA has been evaluated on four legal documents, wherein the question-answer pairs were identified by two experts -- one expert in legal informatics and the other in requirements engineering~\cite{Abualhaija:2022}. In the following, we describe the four documents:

\sectopic{$\bullet$ GDPR} or General Data Protection Regulation (EU) 2016/679 is the European privacy law that harmonises the data protection, privacy and personal data transfer requirements~\cite{GDPR}. The experts identified 36 question-answer pairs from the entire document. The document was partitioned in 301 context spans by the ``Text Preprocessing'' step of Figure~\ref{fig:tool}.

\sectopic{$\bullet$ Directive (EU) 2019/770} is the European directive for regulating the supply of digital content or digital services, and laying down rules for contracts between any trader and consumer of digital content or service~\cite{EU_770}. The experts identified 33 question-answer pairs in this document, and the document was split into 120 context spans.

\sectopic{$\bullet$ Directive (EU) 2019/771} is the European directive concerning the sale of goods~\cite{EU_771}. Directive (EU) 2019/771 complements Directive (EU) 2019/770, as it formalises the contracts on the sale of goods that contain digital elements that require digital content or service. For example, the regulations related to the contracts of the smartphone are covered by Directive (EU) 2019/771, whereas the regulations for operating systems or apps on the smartphone might be covered by Directive (EU) 2019/770. The experts identified 19 question-answer pairs in Directive (EU) 2019/771, and the document text was split into 102 context spans.

\sectopic{$\bullet$ Luxembourg Law of 25 March 2020} is an amendment to numerous existing finance and banking related laws in Luxembourg. The amendment was intended to set due diligence measures for a central electronic data retrieval system related to bank accounts and safe-deposit boxes in Luxembourg, and was a step towards tackling money laundering~\cite{law_2020}. The experts identified 19 question-answer pairs in this legal document. The document text was split into 23 context spans.

To identify the most accurate similarity metric for context span retrieval, we compared BCE (Section~\ref{subsec:contextRetrieval}) with TF-IDF similarity~\cite{Aizawa:03} -- a metric commonly used in the NLP domain. BCE was significantly more accurate than TF-IDF in our experiments. Overall, from the 107 questions over the four documents, BCE retrieves the correct context span for 100 questions (for the top-$5$ spans).
In addition to RoBERTa, we further evaluated three QA models, namely BERT~\cite{Devlin:18}, ALBERT~\cite{Lan:19-albert}, ELECTRA~\cite{Clark:20-electra} for answer extraction (Section~\ref{subsec:answerExtraction}). RoBERTa was deemed the most accurate as it correctly extracted the answers for 97 questions. 

We also analyzed the questions where COREQQA did not highly rank the correct context span (within top-5) or RoBERTa model did not extract the correct answer. Our analysis showed that generic questions, such as the ones formulated for defining or elaborating on a legal concept, were not correctly answered. This is because the legal document (or even a given context span) would usually contain several instances of such legal concept, thus misleading both the context span retrieval and answer extraction steps of COREQQA. We also realized that complicated questions (e.g., with composite conditions) were difficult to answer for COREQQA. 

Last but not least, COREQQA answers questions within reasonable execution time. Thus, in short, our evaluation indicates that COREQQA produces accurate results and is fit for use by requirements engineers in practice. For answering one question from a legal document including an average of 620 sentences, COREQQA requires a total of $\approx$34 seconds.

\section{Usage Scenario}~\label{sec:usabilityScenario}

In this section, we describe an end-to-end example illustrating how our tool can be applied in practice by a requirements engineer. Let \textit{KoopaApp} be a new system under development. KoopaApp is a gym fitness app for maintaining users' workout information and other health-related data. KoopaApp accesses personal information such as the location from other apps on the users' smartphone. 
During the pandemic, such applications often raised concerns about privacy. For example, several health applications were analyzed for privacy-related issues in the RE literature~\cite{Bano:21,Fazzini:22}.  

A requirements engineer (\textit{Daisy}) is in charge of specifying the KoopaApp requirements, including compliance requirements. As an example, we focus only on a subset of compliance requirements related to privacy and data protection. 
Daisy (as is often the case in most software projects) is not very familiar with the privacy regulations, yet she knows well the functionalities and characteristics of the KoopaApp. During the elicitation of requirements, Daisy identifies a set of functionalities that make use of personal data and are thus subject to compliance. Some of these functionalities are related to the security of collected personal data. We assume here that Daisy or her team are aware of the relevant legal documents for their project. Suggesting relevant documents is beyond the scope of COREQQA. Accounting to possible security threats, Daisy poses a question (``What is the procedure for handling a personal data breach?'') using the COREQQA tool on the GDPR~\cite{GDPR}. COREQQA in turn provides the output shown in Figure~\ref{fig:example}.

From the output of COREQQA, Daisy is able to formulate the following compliance requirements (prefixed with the ID $CR$) under the label \textit{Users Data Breach}.\\
\textbf{Notify Users about the Data Breach.} \\
    $CR_1.$ If a data breach is identified on the KoopaApp server, the KoopaApp-NotifyService shall inform the affected users. \\
    $CR_2.$ The KoopaApp-NotifyService shall email the affected users on the registered email address and store the `user informed' response on the server. \\
    $CR_3.$ The KoopaApp-NotifyService shall notify the affected users on the app and store the `read' response on the server.\\    
\textbf{Notify the CIO about the Data Breach.} \\
    $CR_4.$ If a data breach  is identified on the KoopaApp server, the KoopaApp-NotifyService shall send an email to the Chief Information Officer notifying the breach, within 72 hours of its occurrence. 
    
The four compliance requirements fulfill the regulations provided in Figure~\ref{fig:example}.

\section{Conclusion}

We presented COREQQA -- a tool for assisting requirements engineers in better understanding compliance requirements through question-answering based on regulatory or legal documents. COREQQA is developed using a manually created dataset that combines a joint effort of a requirements engineer and a legal expert over four diverse legal documents. The tool is based on recent large-scale language models that are pre-trained for question-answering. Specifically, the tool applies the Sentence BERT cross encoder for retrieving the most relevant text passages from a legal document for a given question. The tool further employs the RoBERTa question-answering model for highlighting the likely answers to the question in the retrieved text passages. 

In future, we plan to conduct a user study to assess how useful COREQQA is in practice.

\begin{acks}
This paper was supported by Central Legislative Service (SCL) -- Government of Luxembourg, the Luxembourg National Research Fund (FNR) under grants PUBLIC2-17/IS/11801776, and by NSERC of Canada under the Discovery and CRC programs.
\end{acks}

\newpage

\bibliographystyle{ACM-Reference-Format}
\balance
\bibliography{paper}

\end{document}